\documentclass[12pt,a4paper]{article}
\usepackage{amsmath,amssymb,amsthm}
\usepackage{authblk}

\usepackage{graphicx}
\usepackage{appendix}
\usepackage{fullpage}
\usepackage{parskip}
\usepackage{setspace}
\usepackage{multirow}
\usepackage{multicol}
\usepackage{bm}
\usepackage{natbib}
\usepackage{enumitem}
\usepackage{thmtools}
\usepackage{color}
\usepackage[dvipsnames]{xcolor}
\usepackage{hyperref}
\definecolor{darkblue}{rgb}{0, 0, 0.5}
\hypersetup{
    colorlinks = true,
    citecolor = darkblue,
    linkcolor = Maroon
}
\usepackage{enumitem}
\usepackage[a4paper,margin=1in]{geometry}
\newcommand{\E}{\operatorname{\mathbb{E}}}
\newcommand{\R}{\mathbb{R}}
\newcommand{\Z}{\mathcal{Z}}
\newcommand{\eqdist}{\overset{d}{=}}

\newtheorem*{remark*}{Remark}

\newtheorem{lem}{Lemma}
\newtheorem{theorem}{Theorem}
\newtheorem{corollary}{Corollary}
\newtheorem{assumption}{Assumption}

\title{\textbf{Kotlarski's lemma for dyadic models}\footnote{We thank two anonymous referees for helpful comments and suggestions.}}

\author[1]{\textsc{Grigory Franguridi}\thanks{Corresponding author. Email: \texttt{franguri@usc.edu}}}
\author[2]{\textsc{Hyungsik Roger Moon}\thanks{Email: \texttt{moonr@usc.edu}}}
\affil[1]{Center for Economic and Social Research, University of Southern California \vspace{1ex}}
\affil[2]{Department of Economics, University of Southern California}

\date{\today}

\begin{document}

\maketitle

\onehalfspacing

\begin{abstract}

    We show how to identify the distributions of the latent components in the two-way dyadic model for bipartite networks $y_{i,\ell}= \alpha_i+\eta_{\ell}+\varepsilon_{i,\ell}$.
    This is achieved by a repeated application of the extension of the classical lemma of \citet{kotlarski1967characterizing} in \citet{evdokimov2012some}.
    We provide two separate sets of assumptions under which all the latent distributions are identified.
    Both rely on some of the latent components being identically distributed.
    
\medskip


\medskip

\end{abstract}

\section{Introduction}

Identifying latent variables from observed data is a central problem in econometrics.
One of the important tools for addressing this problem is a lemma of \citet{kotlarski1967characterizing} and its variants, which provide conditions under which the distribution (characteristic function) of latent variables is identified.

Kotlarski's lemma has been used for identification, estimation, and inference in a variety of economic settings, such as measurement error models \citep{li1998nonparametric,li2002robust,schennach2004nonparametric,kurisu2022uniform}, auctions \citep{li2000conditionally,krasnokutskaya2011identification,grundl2019identification,
andreyanov2022secret}, and models of earning dynamics \citep{bonhomme2010generalized,botosaru2018nonparametric,hu2019semiparametric}.
More recently, Kotlarski's lemma has been used for robust inference in \citet{kato2021robust}.
Finally, generalizations of Kotlarski's lemma exist to the cases of multiple error components or unknown factor loadings \citep{szekely2000identifiability,li2020generalization,lewbel2022kotlarski,
lewbel2024identification}.
For a more complete overview of the variants and uses of Kotlarski's lemma, see, e.g., \citet{schennach2016recent}.

The classical Kotlarski lemma used in most applications assumes repeated measurement with a common latent variable and independent errors, for example,
$y_{i,\ell} = c+\alpha_i + \varepsilon_{i,\ell}$, where $\ell=a,b$ for each $i$. 
In this note, we show how to use the classical Kotlarski lemma to identify the standard two-way dyadic model for bipartite networks,
$y_{i,\ell}=c+\alpha_i+\eta_\ell + \varepsilon_{i,\ell}$. We discuss two cases: the partially connected bipartite network and the fully connected bipartite network.
	
Our main theorem relies on the version of Kotlarski's lemma in \citet{evdokimov2012some}, and hence does not assume that the characteristic functions (CF) of the error components have no zeros, which would rule out many distributions of interest, including all continuous distributions with compact support and many discrete distributions.
Instead, the CFs are allowed to have zeros, as long as they do not overlap with zeros of their first derivatives.

\section{Classical lemma by Kotlarski}

In this section, we briefly discuss the classical lemma by \citet{kotlarski1967characterizing} and its extension in \citet{evdokimov2012some}.
Suppose we observe two repeated noisy measurements $X_1,X_2$ of a variable $M$,
\begin{align*}
    X_1 = M + U_1, \\
    X_2 = M + U_2,
\end{align*}
where $U_1$, $U_2$ are noise variables. Assume that $M,U_1,U_2$ are jointly independent and $\E[U_1]=0$.
The goal is to identify the distributions of $M,U_1,U_2$.

Kotlarski's lemma states that, if the CFs $\phi_M$, $\phi_{U_1}$, and $\phi_{U_2}$ of $M$,$U_1$, and $U_2$, respectively, are nonvanishing everywhere, then these CFs (and hence the distributions) can be recovered from the CF of the observables $X_1,X_2$.

The assumption of everywhere nonvanishing CFs rules out many interesting distributions, such as any nondegenerate distribution with compact support and any discrete distribution with finite support.
\citet{evdokimov2012some} provide an extension of Kotlarski's lemma that relaxes this assumption. 
Specifically, it only requires that the real zeros of $\phi_{U_1}$ and its derivative $\phi_{U_1}'$ are disjoint and that the zeros of $\phi_{U_2}$ form a set of isolated points, see Assumption A and Lemma 1(b) in \citet{evdokimov2012some}.

\section{Application to dyadic models}

\subsection{Case of partially connected network}\label{sec:partial-network}

Consider a bipartite network with two sets of nodes, $\{1, 2\}$ and $\{a, b\}$.
Assume that the node pairs $(1, a),(1, b)$, and $(2,a)$ are linked, so that the associated variables $y_{1,a}, y_{1,b}$ and $y_{2,a}$ are observed.
Notice that we do not assume that nodes $2$ and $b$ are linked so that $y_{2,b}$ may be unobserved.
Consider the dyadic model
\begin{align*}
	y_{1,a} &= \alpha_1 + \eta_a + \varepsilon_{1,a}, \\
	y_{1,b} &= \alpha_1 + \eta_b + \varepsilon_{1,b}, \\
	y_{2,a} &= \alpha_2 + \eta_a + \varepsilon_{2,a},
\end{align*}
where $\alpha_1,\alpha_2,\eta_a,\eta_b$ are unobserved random effects and $\varepsilon_{1,a},\varepsilon_{1,b},$ and $\varepsilon_{2,a}$ are idiosyncratic errors.
We assume that all the latent variables are jointly independent.
We are interested in identifying the distributions of all the latent components $\alpha_1,\alpha_2$, $\eta_a,\eta_b$, $\varepsilon_{1,a}$, $\varepsilon_{1,b}$, and $\varepsilon_{2,a}$ from the distributions of $y_{1,a}$, $y_{1,b}$, and $y_{2,a}$.

Our identification strategy consists of two parts.
First, we identify the distributions of $\alpha_1$, $\eta_a$, and $\varepsilon_{1,a}$.
Then, we provide two sets of restrictions under which the remaining distributions are identified: (i) the equality of the distributions of $\varepsilon_{1,a}$, $\varepsilon_{1,b}$, and $\varepsilon_{2,a}$ (Assumption \ref{as:homogeneity}) and (ii) the equality of the distributions of $\alpha_1$ and $\alpha_2$ and those of $\eta_a$ and $\eta_b$ (Assumption \ref{as:homogeneity-fe}).

Let us now provide the intuition on how the distributions of $\alpha_1$, $\eta_a$, and $\varepsilon_{1,a}$ can be identified by a repeated application of Kotlarski's lemma.
First, for a pair $(1,a), (1,b)$, write
\begin{align*}
    y_{1,a} = \alpha_1 + \eta_a + \varepsilon_{1,a} = M + U_1, \\
    y_{1,b} = \alpha_1 + \eta_b + \varepsilon_{1,b} = M + U_2,
\end{align*}
where $M=\alpha_1$, $U_1=\eta_a + \varepsilon_{1,a}$, and $U_2= \eta_b + \varepsilon_{1,b}$.
By Kotlarski's lemma, the CF $\phi_{\alpha_1}$ is identified.
Similarly, for a pair $(1,a), (2,a)$, write
\begin{align*}
    y_{1,a} = \alpha_1 + \eta_a + \varepsilon_{1,a} = \tilde M + \tilde U_1, \\
    y_{2,a} = \alpha_2 + \eta_a + \varepsilon_{2,a} = \tilde M + \tilde U_2,
\end{align*}
where $\tilde M=\eta_a$, $\tilde U_1=\alpha_1 + \varepsilon_{1,a}$, and $\tilde U_2= \alpha_2 + \varepsilon_{2,a}$.
By Kotlarski's lemma, the CF $\phi_{\eta_a}$ is identified. 
Joint independence of $\alpha_1,\eta_a$, and $\varepsilon_{1,a}$ implies 
\begin{align*}
    \phi_{\varepsilon_{1,a}}(t) = \frac{\phi_{y_{1,a}}(t)}{\phi_{\alpha_1}(t) \phi_{\eta_a}(t)},
\end{align*}
identifying the distribution of $\varepsilon_{1,a}$, and hence the distributions of $\varepsilon_{1,b}$ and $\varepsilon_{2,a}$.
Finally, joint independence of $\alpha_2,\eta_a,$ and $\varepsilon_{2,a}$ implies
\begin{align}
    \phi_{\alpha_2}(t) = \frac{\phi_{y_{2,a}}(t)}{\phi_{\eta_a}(t) \phi_{\varepsilon_{2,a}}(t)}, \label{eq:phi-alpha-2}
\end{align}
and joint independence of $\alpha_1,\eta_b,$ and $\varepsilon_{1,b}$ implies
\begin{align}
    \phi_{\eta_b}(t) = \frac{\phi_{y_{1,b}}(t)}{\phi_{\alpha_1}(t) \phi_{\varepsilon_{1,b}}(t)}, \label{eq:phi-eta-b}
\end{align}
identifying the distributions of the remaining components $\alpha_2$ and $\eta_b$.

We now state the assumptions needed to make the intuition above rigorous.
For a random variable $\zeta$, denote by $\Z_\zeta$ the set of zeros of its CF $\phi_\zeta$ and denote by $\Z_\zeta'$ the set of zeros of the derivative $\phi_\zeta'$ of its CF.

\begin{assumption}\label{as:model}
    \begin{enumerate}[label=(\roman*)]
        \item $\alpha_1,\alpha_2,\eta_a,\eta_b,\varepsilon_{1,a},\varepsilon_{1,b},\varepsilon_{2,a}$ are integrable with zero means.
        \item $\alpha_1,\alpha_2,\eta_a,\eta_b,\varepsilon_{1,a},\varepsilon_{1,b},\varepsilon_{2,a}$ are jointly independent.
    \end{enumerate}
\end{assumption}

\begin{assumption}\label{as:main}
    \begin{enumerate}[label=(\roman*)]
        \item The sets $\Z_{\alpha_1}, \Z_{\eta_a}, \Z_{\varepsilon_{1,a}}$ are pairwise disjoint. \label{alpha-eta-eps}
        \item The sets $\Z_{\alpha_1}$ and $\Z_{\alpha_1}'$ are disjoint. \label{alpha}
        \item The sets $\Z_{\eta_a}$ and $\Z_{\eta_a}'$ are disjoint. \label{eta}
        \item The sets $\Z_{\varepsilon_{1,a}}$ and $\Z_{\varepsilon_{1,a}}'$ are disjoint. \label{eps}
        \item The sets $\Z_{\alpha_2}, \Z_{\eta_b}, \Z_{\varepsilon_{1,b}}, \Z_{\varepsilon_{2,a}}$ consist of isolated points. \label{isolated}
    \end{enumerate}
\end{assumption}

In the identification strategy described above, we apply Kotlarski's lemma in the case when the error terms $U_1,U_2,\tilde U_1,\tilde U_2$ consist of two latent components.
The lemma relies on assumptions about these error terms, which are not primitives of our dyadic model.
Instead, we want to restrict the latent components in a way that would guarantee that the assumptions on the error terms hold.
The following abstract result shows how this can be achieved.

\begin{lem}\label{lemma}
    Let $A=B+C$, where $B$ is independent of $C$. Assume that
    \begin{enumerate}
        \item $\Z_B \cap \Z_C = \varnothing$,
        \item $\Z_B \cap \Z_B' = \varnothing$,
        \item $\Z_C \cap \Z_C' = \varnothing$.
    \end{enumerate}
    Then $\Z_A \cap \Z_A' = \varnothing$.
\end{lem}

\begin{proof}
    Take any $t \in \Z_A$. Then either (i) $t \in \Z_B$ or (ii) $t\in \Z_C$.    
    In the case (i), we have $\phi_A'(t) = \phi_B'(t)\phi_C(t) + \phi_B(t) \phi_C'(t) = \phi_B'(t)\phi_C(t)$.
    By condition 1, $t\notin \Z_C$, and by condition 2, $t \notin \Z_B'$. Therefore, $\phi_A'(t)\neq 0$ and so $t\notin \Z_A'$. Case (ii) is analogous.
\end{proof}

We are now ready to state our adaptation of the main theorem of \citet{evdokimov2012some} to dyadic data.

\begin{theorem}\label{theorem}
Under Assumptions \ref{as:model} and \ref{as:main}, $\phi_{\alpha_1}$ and $\phi_{\eta_a}$ are identified and
\begin{align*}
\phi_{\varepsilon_{1,a}}(s) = \frac{\phi_{y_{1,a}}(s)}{\phi_{\alpha_1}(s)\phi_{\eta_a}(s)}, \,\,\, s\notin \Z_{\alpha_1} \cup \Z_{\eta_a}.
\end{align*}
\end{theorem}

\begin{proof}
    We use the notations $M,U_1,U_2,\tilde M,\tilde U_1,\tilde U_2$ from the discussion above.
Assumptions \ref{as:main}\ref{alpha-eta-eps}, \ref{as:main}\ref{eta}, \ref{as:main}\ref{eps} and Lemma \ref{lemma} imply that the zeros of $\phi_{U_1}$ and $\phi_{U_1}'$ are disjoint.
Assumption \ref{as:main}\ref{isolated} implies that the zeros of $\phi_{U_2}$ are isolated.
Combining with Assumption \ref{as:model} proves Assumption A in \citet{evdokimov2012some}.
Invoking their Lemma 1(b) establishes identification of $\phi_{\alpha_1}$.

Similarly, Assumptions \ref{as:main}\ref{alpha-eta-eps}, \ref{as:main}\ref{alpha}, \ref{as:main}\ref{eps} and Lemma \ref{lemma} imply that the zeros of $\phi_{\tilde U_1}$ and $\phi_{\tilde U_1}'$ are disjoint.
Assumption \ref{as:main}\ref{isolated} implies that the zeros of $\phi_{\tilde U_2}$ are isolated.
Combining with Assumption \ref{as:model} proves Assumption A in \citet{evdokimov2012some}.
Invoking their Lemma 1(b) establishes identification of $\phi_{\eta_a}$.

Finally, independence implies $\phi_{y_{1,a}}(t)=\phi_{\alpha_1}(t) \phi_{\eta_a}(t) \phi_{\varepsilon_{1,a}}(t)$, completing the proof.
\end{proof}

Theorem \ref{theorem} establishes the distributional identification of $\alpha_1$ and $\eta_a$ and also shows that $\phi_{\varepsilon_{1,a}}$ is identified at all points of the real line except for zeros of $\phi_{\alpha_1}$ or $\phi_{\eta_a}$.
We now provide two sets of assumptions under which \emph{all} the latent distributions are identified (which we call \emph{total identification}).
Both sets of assumptions include the following condition.

\begin{assumption}\label{as:isolated}
    The sets $\Z_{\alpha_1}$ and  $\Z_{\eta_a}$ consist of isolated points.
\end{assumption}

Let us show that total identification holds under the distributional homogeneity of the three error terms $\varepsilon_{1,a}, \varepsilon_{1,b}, \varepsilon_{2,a}$.

\begin{assumption}\label{as:homogeneity}
    $\varepsilon_{1,a}\eqdist \varepsilon_{1,b} \eqdist \varepsilon_{2,a}$.
\end{assumption}

\begin{corollary}
    Under Assumptions \ref{as:model}, \ref{as:main}, \ref{as:isolated}, and \ref{as:homogeneity}, the distributions of all the latent variables $\alpha_1,\alpha_2$, $\eta_a,\eta_b$, $\varepsilon_{1,a}$, $\varepsilon_{1,b}$, and $\varepsilon_{2,a}$ are identified.
\end{corollary}

\begin{proof}
    By Assumption \ref{as:isolated}, $\Z_{\alpha_1} \cup \Z_{\eta_a}$ consists of isolated points, and hence the formula for $\phi_{\varepsilon_{1,a}}(s)$ in Theorem \ref{theorem} can be extended to all $s\in\R$ by continuity.
    This identifies the distribution of $\phi_{\varepsilon_{1,a}}$, and, in view of Assumption \ref{as:homogeneity}, the distributions of $\varepsilon_{1,b}$ and $\varepsilon_{2,a}$.
    The formulas \eqref{eq:phi-alpha-2} and \eqref{eq:phi-eta-b} then identify the distributions of $\alpha_2$ and $\eta_b$.
\end{proof}

Next, we show that total identification holds under the distributional homogeneity of the random effects, $\alpha_1, \alpha_2$ and $\eta_a, \eta_b$.

\begin{assumption}\label{as:homogeneity-fe}
    $\alpha_1 \eqdist \alpha_2$ and $\eta_a \eqdist \eta_b$.
\end{assumption}

\begin{corollary}
    Under Assumptions \ref{as:model}, \ref{as:main}, and \ref{as:homogeneity-fe}, the distributions of all the latent variables $\alpha_1,\alpha_2$, $\eta_a,\eta_b$, $\varepsilon_{1,a}$, $\varepsilon_{1,b}$, and $\varepsilon_{2,a}$ are identified.\footnote{Notice that Assumption \ref{as:isolated} is implied by Assumptions \ref{as:main}(v) and \ref{as:homogeneity-fe}.}
\end{corollary}

\begin{proof}
    By Theorem \ref{theorem}, $\phi_{\alpha_1}$ and $\phi_{\eta_a}$ are identified everywhere. By Assumption \ref{as:homogeneity-fe}, $\phi_{\alpha_1}=\phi_{\alpha_2}:=\phi_\alpha$ and $\phi_{\eta_a}=\phi_{\eta_b}:=\phi_\eta$. Finally, joint independence yields
    \begin{align*}
        \phi_{\varepsilon_{1,b}} = \frac{\phi_{y_{1,b}}(s)}{\phi_{\alpha}(s) \phi_{\eta}(s)}, \quad s\notin \Z_{\alpha} \cup \Z_{\eta}, \\
        \phi_{\varepsilon_{2,a}} = \frac{\phi_{y_{2,a}}(s)}{\phi_{\alpha}(s) \phi_{\eta}(s)}, \quad s\notin \Z_{\alpha} \cup \Z_{\eta}.        
    \end{align*}
    By Assumption \ref{as:isolated}, $\Z_{\alpha} \cup \Z_{\eta}$ consists of isolated points, and hence the formulas above can be extended by continuity. This identifies the distributions of $\varepsilon_{1,b}$ and $\varepsilon_{2,a}$.
\end{proof}

\subsection{Case of fully connected network}

When the bipartite network on the node sets $\{1,2\}$ and $\{a,b\}$ is fully connected, i.e., when all the node pairs $(1,a),(1,b),(2,a)$, and $(2,b)$ are linked, the distributions of all the latent variables $\alpha_1,\alpha_2$, $\eta_a,\eta_b$, $\varepsilon_{1,a}$, $\varepsilon_{1,b}$ , $\varepsilon_{2,a}$, and $\varepsilon_{2,b}$ are identified without any assumptions on the distribution homogeneity (cf. Assumptions \ref{as:homogeneity} and \ref{as:homogeneity-fe}).\footnote{For this, Assumption \ref{as:main} has to be extended to include the conditions on $\varepsilon_{2,b}$.}
To see that, notice that applying Kotlarski's lemma to the pair $y_{i,a}, y_{i,b}$ identifies the distribution of $\alpha_i$, $i=1,2$.
Then applying the lemma to the pair $y_{1,c}, y_{2,c}$ identifies the distribution of $\eta_c$, $c=a,b$.
Finally, the distribution of $\varepsilon_{i,c}$ is identified via
\begin{align*}
    \phi_{\varepsilon_{i,c}}(t) = \frac{\phi_{y_{i,c}}(t)}{\phi_{\alpha_i}(t)\phi_{\eta_c}(t)}, \quad i=1,2, \,\, c=a,b.
\end{align*}
Formulating rigorous conditions under which this identification strategy is valid can be done along the lines of Section \ref{sec:partial-network}.

\section{Conclusion}

We show how the classical lemma of Kotlarski can be employed to identify distributions of latent components in dyadic models for bipartite networks with two-way random effects.
When the bipartite graph is partially linked, we provide two sets of assumptions under which all the latent distributions are identified.
The first set of assumptions restricts the errors to be identically distributed.
The second set of assumptions restricts the random effects to be identically distributed.
Our identification result may be applicable to dyadic regressions with random or correlated effects, as discussed in Section 2.3 of \citet{bonhomme2020econometric}.
It can also be used to develop an estimation procedure for this class of models. This is beyond the scope of the current paper, and we defer this agenda to future work.

\bibliographystyle{ecta}
\bibliography{references}

@article{evdokimov2012some,
  title={Some extensions of a lemma of Kotlarski},
  author={Evdokimov, Kirill and White, Halbert},
  journal={Econometric Theory},
  volume={28},
  number={4},
  pages={925--932},
  year={2012},
  publisher={Cambridge University Press}
}

@article{li1998nonparametric,
  title={Nonparametric estimation of the measurement error model using multiple indicators},
  author={Li, Tong and Vuong, Quang},
  journal={Journal of Multivariate Analysis},
  volume={65},
  number={2},
  pages={139--165},
  year={1998},
  publisher={Elsevier}
}

@article{kotlarski1967characterizing,
  title={On characterizing the gamma and the normal distribution},
  author={Kotlarski, Ignacy},
  journal={Pacific Journal of Mathematics},
  volume={20},
  number={1},
  pages={69--76},
  year={1967},
  publisher={Mathematical Sciences Publishers}
}

@article{schennach2016recent,
  title={Recent advances in the measurement error literature},
  author={Schennach, Susanne M},
  journal={Annual Review of Economics},
  volume={8},
  number={1},
  pages={341--377},
  year={2016},
  publisher={Annual Reviews}
}

@article{schennach2004nonparametric,
  title={Nonparametric regression in the presence of measurement error},
  author={Schennach, Susanne M},
  journal={Econometric Theory},
  volume={20},
  number={6},
  pages={1046--1093},
  year={2004},
  publisher={Cambridge University Press}
}

@article{li2002robust,
  title={Robust and consistent estimation of nonlinear errors-in-variables models},
  author={Li, Tong},
  journal={Journal of Econometrics},
  volume={110},
  number={1},
  pages={1--26},
  year={2002},
  publisher={Elsevier}
}

@article{szekely2000identifiability,
  title={Identifiability of distributions of independent random variables by linear combinations and moments},
  author={Sz{\'e}kely, GJ and Rao, CR},
  journal={Sankhy{\=a}: The Indian Journal of Statistics, Series A},
  pages={193--202},
  year={2000},
  publisher={JSTOR}
}

@article{li2020generalization,
  title={A generalization of Lemma 1 in Kotlarski (1967)},
  author={Li, Siran and Zheng, Xunjie},
  journal={Statistics \& Probability Letters},
  volume={165},
  pages={108814},
  year={2020},
  publisher={Elsevier}
}

@article{kato2021robust,
  title={Robust inference in deconvolution},
  author={Kato, Kengo and Sasaki, Yuya and Ura, Takuya},
  journal={Quantitative Economics},
  volume={12},
  number={1},
  pages={109--142},
  year={2021},
  publisher={Wiley Online Library}
}

@article{li2000conditionally,
  title={Conditionally independent private information in OCS wildcat auctions},
  author={Li, Tong and Perrigne, Isabelle and Vuong, Quang},
  journal={Journal of econometrics},
  volume={98},
  number={1},
  pages={129--161},
  year={2000},
  publisher={Elsevier}
}

@article{krasnokutskaya2011identification,
  title={Identification and estimation of auction models with unobserved heterogeneity},
  author={Krasnokutskaya, Elena},
  journal={The Review of Economic Studies},
  volume={78},
  number={1},
  pages={293--327},
  year={2011},
  publisher={Oxford University Press}
}

@article{grundl2019identification,
  title={Identification and estimation of risk aversion in first-price auctions with unobserved auction heterogeneity},
  author={Grundl, Serafin and Zhu, Yu},
  journal={Journal of Econometrics},
  volume={210},
  number={2},
  pages={363--378},
  year={2019},
  publisher={Elsevier}
}

@article{kurisu2022uniform,
  title={On the uniform convergence of deconvolution estimators from repeated measurements},
  author={Kurisu, Daisuke and Otsu, Taisuke},
  journal={Econometric Theory},
  volume={38},
  number={1},
  pages={172--193},
  year={2022},
  publisher={Cambridge University Press}
}

@article{andreyanov2022secret,
  title={Secret reserve prices by uninformed sellers},
  author={Andreyanov, Pasha and Caoui, El Hadi},
  journal={Quantitative Economics},
  volume={13},
  number={3},
  pages={1203--1256},
  year={2022},
  publisher={Wiley Online Library}
}

@article{bonhomme2010generalized,
  title={Generalized non-parametric deconvolution with an application to earnings dynamics},
  author={Bonhomme, St{\'e}phane and Robin, Jean-Marc},
  journal={The Review of Economic Studies},
  volume={77},
  number={2},
  pages={491--533},
  year={2010},
  publisher={Wiley-Blackwell}
}

@article{botosaru2018nonparametric,
  title={Nonparametric heteroskedasticity in persistent panel processes: An application to earnings dynamics},
  author={Botosaru, Irene and Sasaki, Yuya},
  journal={Journal of Econometrics},
  volume={203},
  number={2},
  pages={283--296},
  year={2018},
  publisher={Elsevier}
}

@article{hu2019semiparametric,
  title={Semiparametric estimation of the canonical permanent-transitory model of earnings dynamics},
  author={Hu, Yingyao and Moffitt, Robert and Sasaki, Yuya},
  journal={Quantitative Economics},
  volume={10},
  number={4},
  pages={1495--1536},
  year={2019},
  publisher={Wiley Online Library}
}

@article{lewbel2022kotlarski,
  title={Kotlarski with a factor loading},
  author={Lewbel, Arthur},
  journal={Journal of Econometrics},
  volume={229},
  number={1},
  pages={176--179},
  year={2022},
  publisher={Elsevier}
}

@article{lewbel2024identification,
  title={Identification of a triangular two equation system without instruments},
  author={Lewbel, Arthur and Schennach, Susanne M and Zhang, Linqi},
  journal={Journal of Business \& Economic Statistics},
  volume={42},
  number={1},
  pages={14--25},
  year={2024},
  publisher={Taylor \& Francis}
}

@incollection{bonhomme2020econometric,
  title={Econometric analysis of bipartite networks},
  author={Bonhomme, St{\'e}phane},
  booktitle={The econometric analysis of network data},
  pages={83--121},
  year={2020},
  publisher={Elsevier}
}

\end{document}